\documentclass[12pt]{elsart} 
\usepackage{amsmath}
\usepackage{graphicx}
\usepackage[T1]{fontenc}
\usepackage{moreverb}
\usepackage{subfigure}
\usepackage{xspace}
\usepackage{float}

\newcommand{\XANES}{{\sc xanes}}
\newcommand{\EELS}{{\sc eels}}

\newcommand{\TEM}{{\sc tem}}

\newcommand{\MDFF}{{\sc mdff}}
\newcommand{\DFF}{{\sc dff}}
\newcommand{\ELMCD}{{\sc emcd}}
\newcommand{\EMCD}{{\sc emcd}}

\newcommand{\XMCD}{{\sc xmcd}}

\begin{document}  
\begin{frontmatter}

\title{Magnetic circular dichroism in EELS: Towards 10 nm resolution}

\author[IFP,USTEM]{Peter Schattschneider}
\author[IFP]{C\'{e}cile H\'{e}bert}
\author[IFP]{Stefano Rubino} 
\author[USTEM]{Michael St\"oger-Pollach}
\author[IOP]{Jan Rusz}
\author[IOP]{ Pavel Nov\'{a}k}
\address[IFP]{Institute for Solid State Physics, Vienna University of Technology, A-1040 Vienna, Austria}
\address[USTEM]{University Service Centre for Electron Microscopy, Vienna University of Technology, A-1040 Vienna, Austria}
\address[IOP]{Institute of Physics, Academy of Sciences of the Czech Republic, Cukrovarnick\'{a} 10, 16253 Prague 6, Czech Republic}
\date{\today}

\begin{abstract}
We describe a new experimental setup for the detection of magnetic circular dichroism with fast electrons (\EMCD{}). As compared to earlier findings the signal is an order of magnitude higher, while the probed area could be significantly reduced, allowing a spatial resolution of the order of 30~nm.  
A simplified analysis of the experimental results is based on the decomposition 
of the Mixed Dynamic Form Factor $S(\vec q, \, \vec q',E)$ into a real part
related to the scalar product and an imaginary part related to the vector 
product of  the scattering vectors ${\vec q}$ and ${\vec q'}$. Following the 
recent detection of chiral electronic transitions in the electron microscope 
the present experiment is a crucial demonstration of the potential of \EMCD{} for nanoscale investigations. 
\end{abstract}

\end{frontmatter}

\section{Introduction}
The observation of circular dichroism with electron probes has been considered 
impossible except with spin polarized electron probes. In 2003 it was suggested 
that this was not the case~\cite{HebertUM03}. The magnetic transitions that give
rise to X-ray magnetic circular dichroism (\XMCD{}) contribute to the imaginary
part of the Mixed Dynamic Form Factor (\MDFF{})~\cite{KohlUM85} for inelastic 
electron scattering. Since this quantity can be measured in the transmission 
electron microscope (\TEM{}) under particular scattering conditions, we predicted
that detection of \XMCD{} should be feasible in the \TEM{}. We called the predicted 
effect Energy-Loss Magnetic Chiral Dichroism (\ELMCD{}).

In the first conclusive experimental demonstration of \EMCD{} on Fe~\cite{Nature2006}
it was discovered that the effect is smaller than \XMCD{}. In that experiment the dichroic signal was close to the noise threshold in the then chosen geometry and the area
of investigation was approximately 100~nm in radius.

Here we present a new experimental setup that enhances the count rate by an order of magnitude and 
reduces the probed area by another factor of five, thus opening the way to applications of \EMCD{} on the nanometric scale.

The \MDFF{} $S({\vec q}, {\vec q}',E)$ is the essential quantity describing 
\EMCD{}. It has been used in the description of interference of inelastically 
scattered electrons ({\it e.~g.}~\cite{KohlUM85,SchattM99,SchattschneiderElspec2005}). 

\section{The Mixed Dynamic Form Factor (MDFF)}
The semi-relativistic double differential scattering cross section for inelastic
electron scattering (DDSCS) in the plane wave Born approximation
is~\cite{SchattschneiderPRB05}
\begin{equation}
\frac{\partial^2\sigma}{\partial E\partial\Omega}= \frac{4 \gamma^2}{a_0^2 \, q^4}  \frac{k_f}{k_i}S({\vec q},E)
\label{DDSCS}
\end{equation}
where $a_0$ is the Bohr radius, $k_i$ $(k_f)$ is the wave number of the 
incident (outgoing) probe electron, ${\vec q} = {\vec k}_i - {\vec k}_f $ is the
wave vector transfer in the interaction and $E$ the energy loss. 
$S({\vec q},E)$ is the dynamic form factor (\DFF{}). 

Interference between inelastically 
scattered electrons in the diffraction pattern will occur when the probe 
electron consists of two or more mutually coherent plane waves~\cite{SchattM99,SchattPRB99}. Experimentally, this can be realized by a 
biprism~\cite{Lichte2000} or by any other beam splitter. It was shown 
experimentally that the crystal itself can be used as a beam splitter for 
inelastic electron scattering~\cite{NelhiebelPRL00}. In the crystal the probe 
electron is a superposition of Bloch waves which, in turn, are coherent 
superpositions of plane waves defined by the allowed Bragg reflections.
 
 For the sake of clarity, we consider the simplest case here, namely the 
superposition of two plane waves with complex amplitudes $A_{1, \, 2}$, respectively. 
Technically, this situation is approximated in electron diffraction by the 
two-beam case, the most important plane waves being the incident one and a single 
Bragg scattered wave.

The DDSCS is then~\cite{KohlUM85}
\begin{equation}
\frac{\partial^2\sigma}{\partial E\partial\Omega}= \frac{4 \gamma^2}{a_0^2 }  \frac{k_f}{k_i}\big(|A_1|^2 \frac{S({\vec q},E)}{q^4} + |A_2|^2 \frac{S({\vec q}\,',E)}{q'^4} + 2 \Re[ A_1 A_2^\ast \frac{S({\vec q}, {\vec q}\,',E)}{q^2q'^2}]\big).
\label{DDSCSinter}
\end{equation}
Here,
${\vec q} = {\vec k}_i - {\vec k}_f $, ${\vec q}\,' = {{\vec k}'}_i - {\vec k}_f $ are 
the wave vector transfers from the two incident plane waves ${\vec k}_i, \, {{\vec k}'}_i$ to 
${\vec k}_f$. Since i) the $2p$ $\to$ $3d$ transitions are the dominant ones due to the 
shape of the density of states and ii) due to the localized character of $<i|$, 
the matrix elements $\langle i |e^{i{\vec q}  {\vec  R}} |f \rangle$ are mostly determined by an area within small $R$ values (compared to lattice parameters), we can use the dipole approximation for the \MDFF{}~\cite{SchattM99}
\begin{equation}
S({\vec q},{\vec q}\,',E)_{\text{dip}}=\sum_{if}  \langle i |{{\vec q}  {\vec  R}} |f \rangle \langle f |{{\vec q}\, '   {\vec  R}} |i  \rangle \delta(E+E_i - E_f).
\label{MDFF}
\end{equation}
 $ {\vec R}$ is the 3-vector operator $(r_1, r_2, r_3)$ of the one-electron 
scatterer with initial and final wave functions $|i \rangle$, $|f \rangle$.
 
 Eq.~\ref{DDSCSinter} consists of two direct terms, each resembling the angular 
scattering distributions centered at the incident and the Bragg scattered plane wave directions, and an interference term. Eq.~\ref{DDSCSinter} is formally equivalent to the expression for intensity in the double slit experiment. It should be noted that the diagonal element of the \MDFF{} is the \DFF{}, $S({\vec q},{\vec q},E)=S({\vec q},E)$.

The \MDFF{} describes the mutual coherence of transitions with energy transfer 
$E$ and momentum transfer $\hbar{\vec q}, \hbar{\vec q}'$~\cite{SchattM99} 
(Two different momentum transfers can occur in one transition with finite 
probability when the incident or the outgoing electron is not a single plane wave. In 
the present case  $ A_1 |{\vec k_i} \rangle + A_2 |{\vec k'}_i \rangle$ is such 
a basis function, and the measurement collapses the probe electron into 
$|{\vec k}_f \rangle$~\footnote{It should be noted that
collapsing the probe electron into $|{\vec k}_f \rangle$ does not
exclude the possibility of interference between outgoing beams with
different wave vectors; any outgoing beam that is Bragg scattered to $|{\vec k}_f \rangle$ before leaving the crystal can produce interference detectable by the setup described here.}).

With the  matrix elements
\begin{equation}
r_{jk}=\sum_{if}  \langle i |r_j |f \rangle \langle f |r_k |i  \rangle \delta(E+E_i - E_f).
\label{rjk}
\end{equation}
of the transition matrix ${  \hat R}=\{r_{jk} \}$
the \MDFF{}, eq.~\ref{MDFF} can be written as
\begin{equation}
S({\vec q},{\vec q}\,',E)_{dip}={\vec q }{  \hat R} {\vec q}\,' \, .
\label{Sdip} 
\end{equation}

For isotropic systems the transition matrix degenerates to a quantity
proportional to the unity matrix~\cite{KohlUM85}. This case was discussed
in the context of ionisation fine structure and dynamical diffraction ~\cite{NelhiebelPM99,SchattPRB99}. 

Anisotropy can be induced by a lattice of lower than cubic symmetry, or by 
magnetic fields. These can be internal or external, then speaking of natural 
or magnetic dichroism~\cite{Lovesey}. It is well known that with photon 
scattering linear as well as circular dichroism can be measured. This 
technique is largely applied with external magnetic fields. Linear magnetic 
dichroism shows up as an uniaxial anisotropy and can be measured with angle 
resolved inelastic electron scattering, tuning $\vec q$ parallel or 
perpendicular to the anisotropy axis. This is equivalent \cite{HitchcockJJAP93} to the tuning of 
linear polarization of the photon in \XANES{} experiments. 

It has been thought that circular magnetic dichroism cannot be detected with 
electrons except with spin polarized ones. But let us recall that in \XANES{}
the photon does not couple directly to the spin of electrons but to the angular 
momentum of the excited atom, and the effect becomes visible by the spin-orbit 
coupling~\cite{Lovesey}. So there is no reason that spin polarized electrons 
are needed for detection of circular dichroism in electron energy loss 
spectrometry (\EELS{}). Rather, in the inelastic electron interaction that is 
equivalent to an \XMCD{} experiment, the virtual photon that is exchanged must 
be circularly polarized.

The \MDFF{}, eq.~\ref{Sdip} can be written in a different form when we specify the magnetic field 
direction
 as the positive $r_3$ axis~\footnote{In the TEM this is usually also the optical axis.} and write 
${\vec q} =({\vec q}_\perp, q_3)$. A little algebra shows that 
\begin{equation}
S({\vec q},{\vec q}\,',E)_{dip}= \frac{1}{2} ( {\rm r}_{++} + {\rm r}_{--} ) \, {\vec q}_\perp \cdot {\vec q}_\perp\,'+ {\rm r}_{00} \, q_3 q'_3 + \frac{i}{2} ( {\rm r}_{++} -  {\rm r}_{--}) \, ({\vec q}_\perp {\times} \,{\vec q}_\perp\,') \cdot {\vec e_3}
\label{Sfinal}
\end{equation}
where ${\vec e_3}$ is the unit vector in direction of the $r_3$ axis, and we have used the transition matrix elements in terms of the spherical components $R_{+, \, -, \,0}$ of the 3-space operator 
\begin{equation}
{\rm r}_{++}=\sum_{if}  \langle i |R_+ |f \rangle \langle f |R_+ |i  \rangle \delta(E+E_i - E_f).
\label{Rij}
\end{equation}
and similar for all other combinations. They relate to the Cartesian components by the transformation rules for vector spherical harmonics \cite{HannonPRL1988}.
All matrix elements and vector components are real in eq.~\ref{Sfinal}. In this 
form we have separated the \MDFF{} into a real component proportional to the 
scalar product of the wave vector transfers ${\vec q}_\perp, \,{\vec q}\,'_\perp$ 
and an imaginary part proportional to their vector product. This structure is 
equivalent to that of the polarization tensor used in \XMCD{}, which decomposes 
into a scalar part (uneffective in dichroic experiments), a second-rank irreducible part detectable by linear dichroism and a pseudovector part
sensitive to magnetic moments \cite{Lovesey,Altarelli} and reference
therein. 
 
This form of the MDFF allows a description of the scattering geometry for \EMCD{} detection in the \TEM{}. 
In passing we note that the imaginary part vanishes if the magnetic transitions ($\Delta m = \pm 1$) are degenerate.
Only when the presence of a magnetic field in $r_3$ direction lifts the $m$-degeneracy will we see an effect. 
For a transition with fixed energy loss $E$ the operators $R_+$ and 
$R_-$ will then contribute with different oscillator strengths, and the \MDFF{} eq.~\ref{Sfinal} 
will acquire an imaginary part. Its sign depends on which transitions are allowed 
by the selection rules. 

The imaginary part can be interpreted as the difference in probability to 
change the magnetic quantum number by $\pm 1$. It thus describes the 
difference in response of the system to left- respectively right-handed 
circularly polarized electromagnetic fields~\footnote{An imaginary part of the 
\MDFF{} signifies that time inversion symmetry is broken. In fact this symmetry 
breaking relates to the angular momentum operator. Under time inversion its 
direction is reversed. In the presence of a magnetic field this is no longer 
a symmetry operation.}.

The scattering vector ${\vec q}_\perp$ in the diffraction plane is 
perpendicular to the magnetic field vector. We assumed already that the 
magnetic moments of the scatterer are aligned parallel to the optical 
axis $r_3$ in the strong magnetic field ($\approx$ 2 T) of the objective lens of the 
microscope. We can now evaluate the DDSCS for specimens showing magnetic 
circular dichroism. 

If in eq.~\ref{DDSCSinter} we write the phase shift $\phi$ explicitly as $A_1 A_2^\ast = |A_1||A_2| \cdot e^{-i\phi}$, inspection of eq.~\ref{Sfinal} then reveals that a 
phase shift $\phi \neq n \pi$ between the two incident plane waves is 
needed in order to activate the imaginary part of the \MDFF{}. A phase shift of $\pm \pi/2$ 
is recommended since in this case the real part of the \MDFF{} disappears
in eq.~\ref{DDSCSinter}, and only the imaginary part survives.
In a two-beam case with such a phase shift the pseudovector part contributes 
with its full magnitude and gives rise to an asymmetry in the scattering cross section of a 
magnetic transition such as the L edges of the ferromagnetic d-metals.

\section{Experiments and simulations}
\begin{figure}
\includegraphics[width=0.48\textwidth]{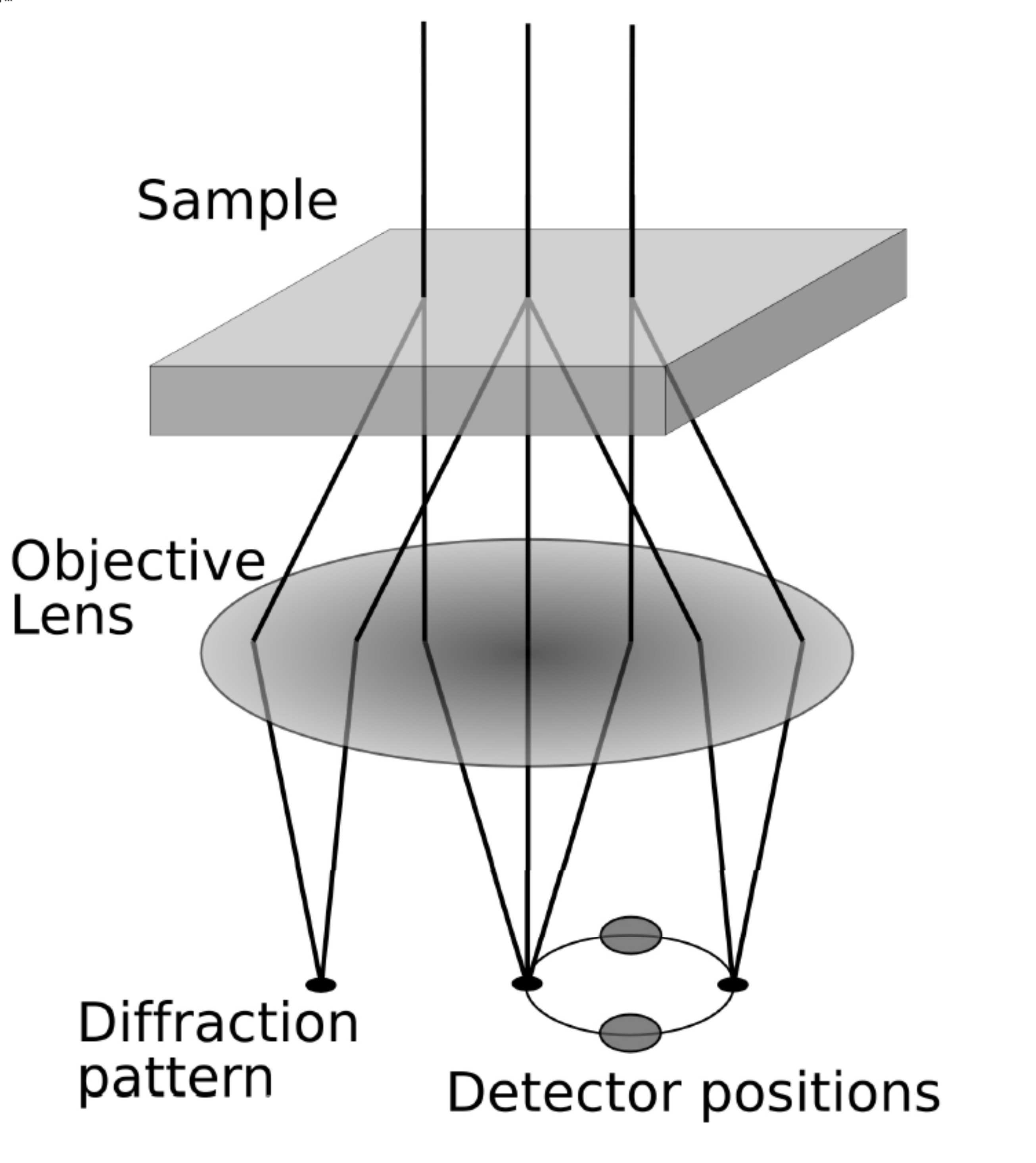}
\includegraphics[width=0.48\textwidth]{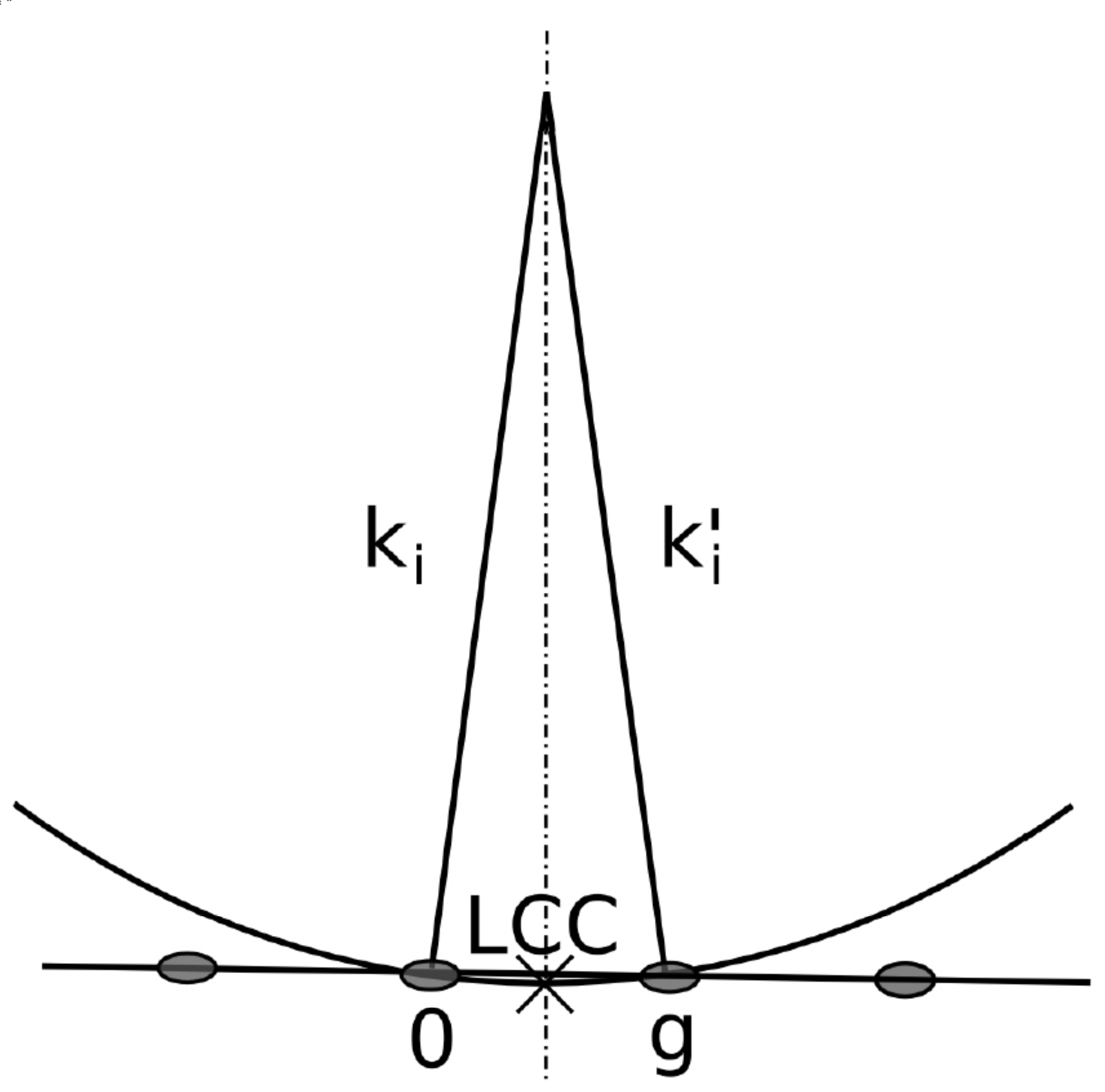}
\caption{Left: the electron beam creates a diffraction pattern in the focal plane of the objective lens. Placing the SEA in the diffraction plane will select one
direction for the outgoing electron beam. The red circles indicate positions A and B of our experimental setup. Right: in reciprocal space the
Ewald sphere is defined by the incident wave vector $\bf k_i$. If a
reciprocal lattice vector  (g) lies on the surface of the sphere, the Bragg spot will be strongly excited (as is the case for $\bf k'_i$ in the figure). The projection of the sphere on the diffraction plane is called Laue Circle Centre (LCC, equal to g/2 in the figure).}
\label{diff}
\end{figure}

\begin{figure}
	\begin{center}
		\includegraphics[scale=0.9]{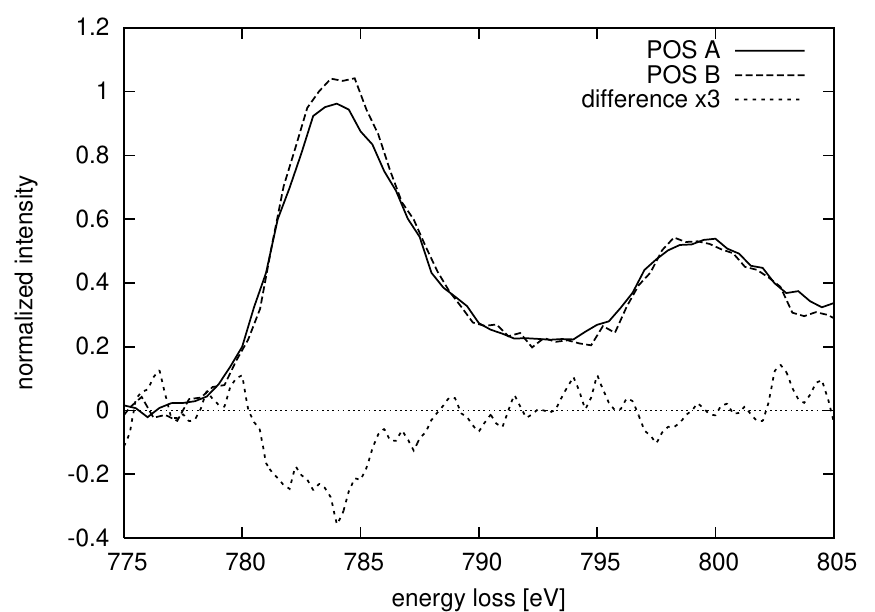}
		\caption{\label{Cospectraold} EELS spectra of the Co $L_{2,3}$ edge showing EMCD using the detector shift method.}
	\end{center}
	
\end{figure}

We performed our experiment on a Co single crystal electropolished sample with a FEI Tecnai F20-FEGTEM S-Twin operating at 200 keV and equipped with a Gatan imaging filter (GIF). The dichroic signal is obtained by first tilting out of the [0001] zone axis to a two-beam case where only the $0000$ and $10\bar 10$ reflections are strongly excited. Following the procedure illustrated in \cite{Nature2006}, a selected area aperture (SAA) is used to delimit a region of about 100 nm radius and 18~$\pm$~3~nm thickness. The corresponding diffraction pattern is then projected onto the 2 mm spectrometer entrance aperture (SEA). 
Drawing a circle with a diameter of G with the 0- and G-reflections on the
left and right side respectively, the strongest dichroic signal can be
expected at the top and bottom points A and B of this Thales circle (fig.~\ref{diff}) where the scalar product ${\vec q} \cdot {\vec q}\,'$ is zero and the pseudovector ${\vec q} \times {\vec q}\,'$ maximises the imaginary part of the \MDFF{}.
The camera lenght and projection coils are then adjusted so that the SEA will be centered at these two points (first A, then B) in the Thales circle and determine a collection angle of 2-4 mrad. Two spectra are then acquired sequentially (fig.~\ref{Cospectraold}). With an acquisition time of 60 sec per spectrum and an energy dispersion of 0.5 eV/channel the intensity at the $L_3$ peak was about 2,500 counts (after background removal).

\begin{figure}
	\begin{center}
		\includegraphics[scale=0.5]{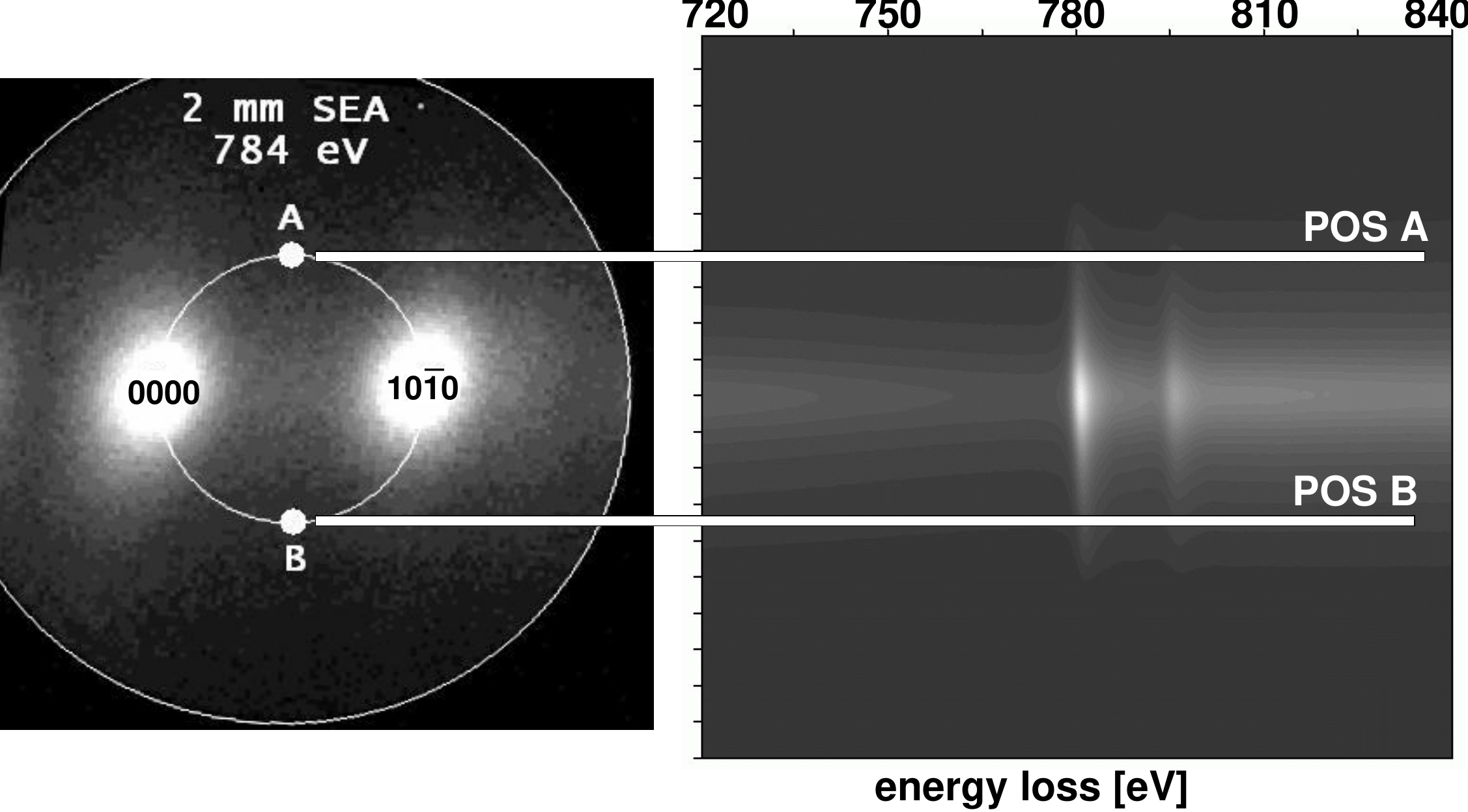}
		\caption{ \label{lacbed} Left: Energy filtered 
		diffraction pattern at 784~eV energy loss using a slit width 
		of 20~eV. The sample was oriented in the two-beam case, 
		capturing both reflections within the 2~mm SEA. 
		Right: q-E diagram of the Co L$_{2,3}$ edge in chiral 
		conditions (for details see text).}
	\end{center}
	
\end{figure}

In order to improve the signal-to-noise ratio a new experimental setup was devised. As detailed above, the sample is tilted out of the [0001] zone axis to a two-beam case where only the $0000$ and $10\bar 10$ reflections are observed in the the energy filtered diffraction pattern, which is then projected onto the spectrometer entrance aperture (SEA). Using a rotational sample holder, the reciprocal lattice vector $\vec{G}$ is then aligned parallel to the energy dispersive axis of the CCD camera, so that a q-E diagram can be recorded as depicted in fig.~\ref{lacbed}. The quadrupoles of the energy filter collapse (integrating the signal in the $q_x$ dimension) the circular area to a line in $q_y$ when the system is switched to spectroscopy mode. This first modification allows us to record not only both spectra A and B with a single acquisition, but the entire range of spectra with different $q_y$ values comprised within the 2 mm SEA. It should be noted however that the integration area in the $q_x$ dimension is different for every value of $|q_y|$.

\begin{figure}
	\begin{center}
		\includegraphics[scale=0.5]{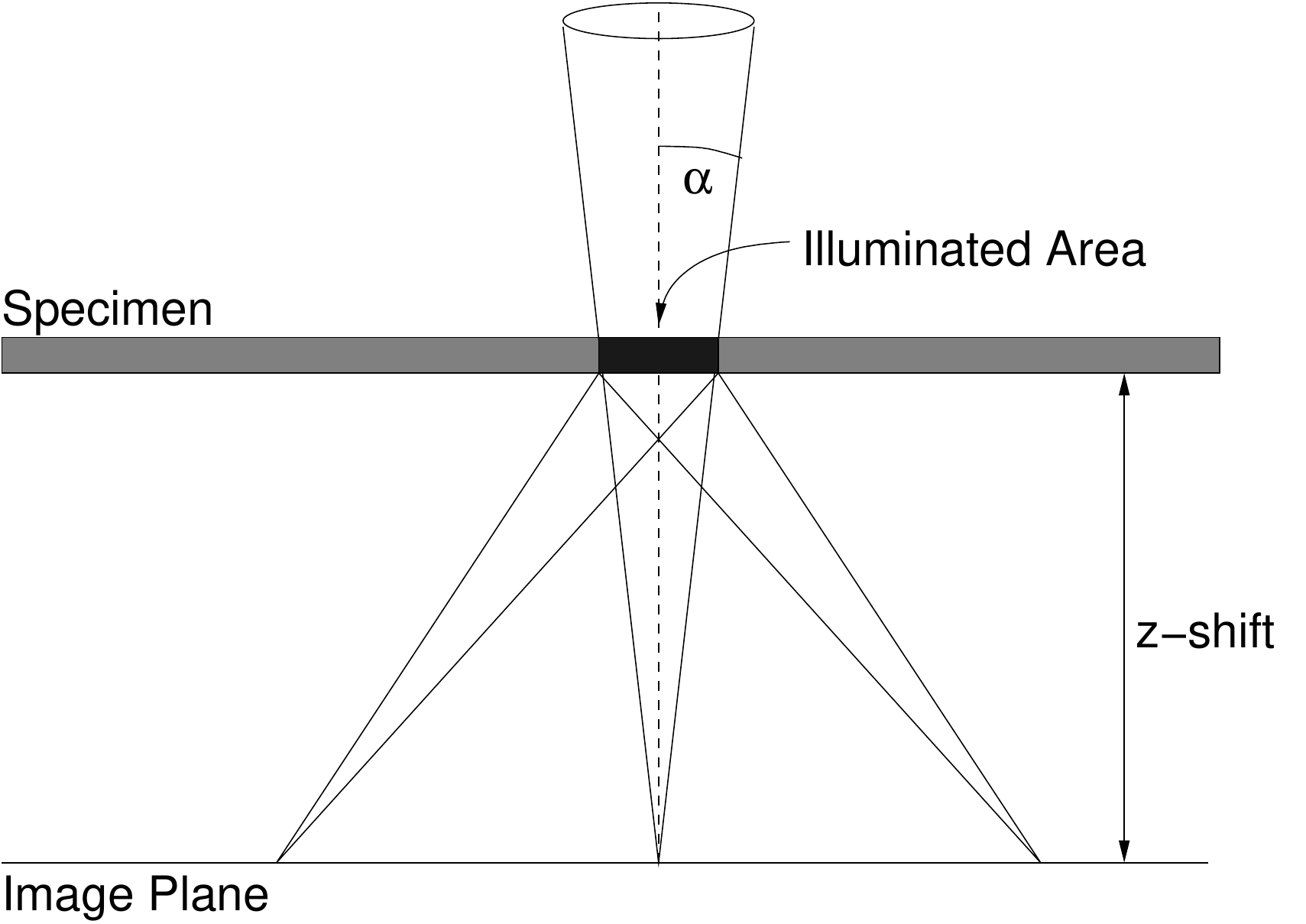}
		\caption{ \label{fig0-dichro0} Geometry of the experiment: When the specimen is shifted upward from the eucentric position (Image Plane)  by $z$ the illuminated area is a disk with radius $\alpha z$.  Bragg scattering will cause a diffraction pattern to appear in the Image Plane. 
}
	\end{center}
	
\end{figure}

A second modification consists in a different method \cite{MidgleyUM99} to obtain a spot like inelastic diffraction pattern: the beam (with a convergence semi-angle of $\alpha = 2$~mrad) is focussed onto a 18~$\pm$~3~nm thick area of the Co specimen which is then shifted upwards from the eucentric position by $z = 9.25 \, \mu$m. The diameter of the illuminated area is  
$d = 2\alpha z=37$~nm, accurate to 5\%, in the present experiment (fig.~\ref{fig0-dichro0}).

Preliminary experiments show that with smaller $z$-shifts the illuminated area can be reduced to less than 10 nm radius, at the moment with untenable distortions of the diffraction pattern. With a C$_\text{s}$ corrector or a monochromator a spatial resolution of 10~nm or less should be attainable. 

\begin{figure}
	\begin{center}
		\includegraphics[scale=0.9]{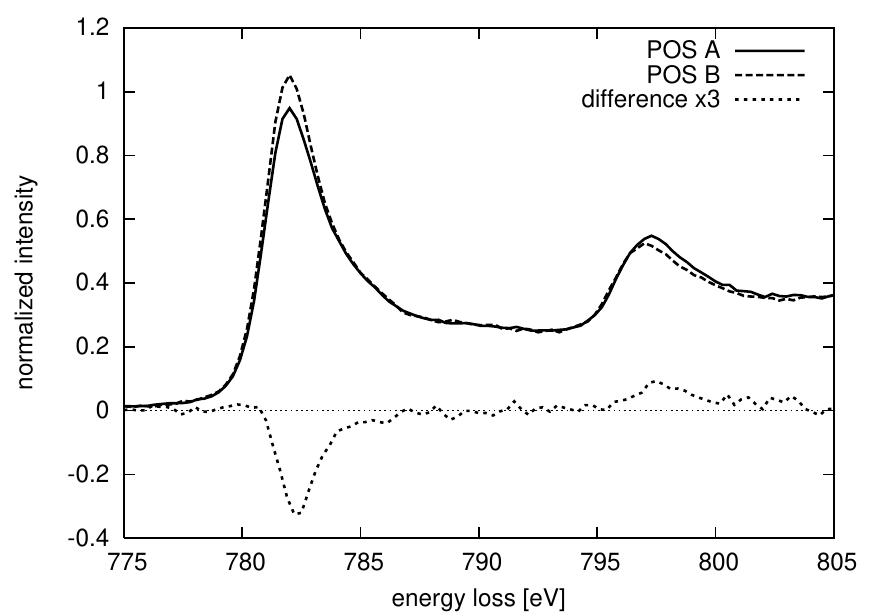}
		\caption{\label{Cospectra} EELS spectra extracted from the 
		q-E diagram (fig. \ref{lacbed}) by line traces parallel to the energy axis and 
		passing by the positions A and B. The difference is the 
		dichroic signal	stemming from $\Im(S(q,q',E))$.}
	\end{center}
	
\end{figure}

Similarly to \XMCD{} we define $\Delta \sigma$ as the difference between spectra with opposite helicity and $\bar \sigma$ as their average. The dichroic signal is then the relative 
difference $\Delta \sigma / \bar \sigma$ in the scattering cross section when 
the sign of the pseudovector part changes~\footnote{Another common definition of the dichroic signal is the ratio between the difference and the sum of spectra with different helicity, which is a factor of 2 smaller than the one used here.}. This is obtained by tracing the 
spectral intensity at points A and B in fig.~\ref{lacbed}, and taking their difference, as shown in fig.~\ref{Cospectra}. With an acquisition time of 15 sec and an energy dispersion of 0.3 eV/channel the intensity at the $L_3$ peak was about 13,500 counts (after background removal).

The modified scattering geometry provides a count rate per eV
which is an order of magnitude higher than the one achieved in the previous configuration~\cite{Nature2006}, thus improving significantly the signal to 
noise ratio. This is essentially caused by the fact that when no SAA is used and the beam is focussed only on the area of interest, all the electrons emitted from the gun contribute to the detected signal. In the formerly used geometry a nearly parallel incident bundle illuminated a large area of the sample of which only a small fraction could be used. This effectively reduced the intensity by which the area of interest is 
illuminated, {\it i.e.} a large part of the incident electrons did not contribute to the signal. 
The increase in the count rate per eV allows us to reduce the acquisition time, thus limiting the effects of beam instability, specimen and energy drift.
The shorter acquisition time, combined with the finer energy dispersion, improves the energy resolution with which the $L_{2,3}$ edges are recorded. In the older setup the $L_3$ has a FWHM of 7 eV (fig.~\ref{Cospectraold}), compared to the 3.6 eV achieved with the new method (fig.~\ref{Cospectra}). 

\begin{figure}
	\begin{center}
		\includegraphics[scale=0.9]{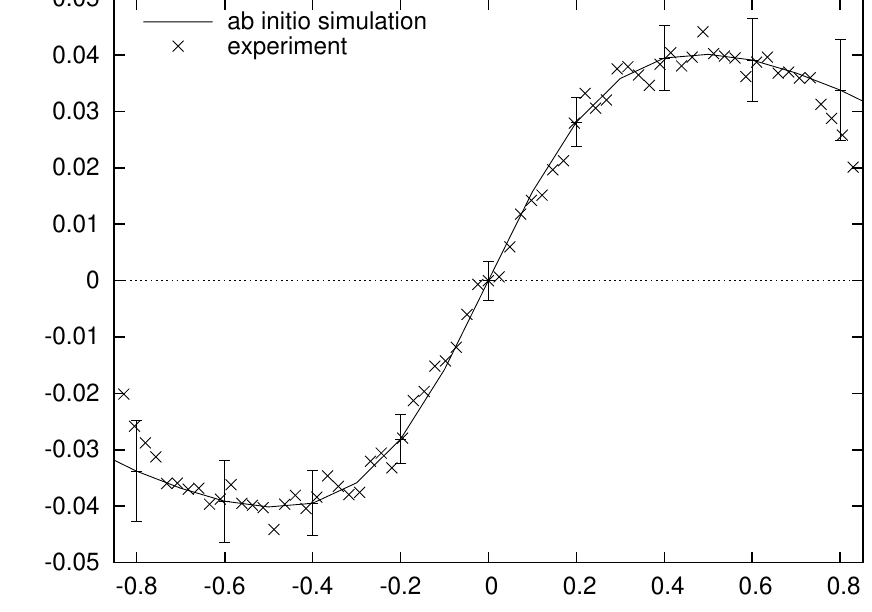}
		\caption{\label{intqyline} Dichroic signal at the Cobalt  $L_3$ edge as function of the scattering angle $q$ (in unit of G) in the direction perpendicular to the Bragg scattering vector G. Positions A and B (see fig.~\ref{lacbed}) correspond to points q/G = 0.5 and -0,5 respectively. Comparison between experimental data (triangles) and {\it ab initio} simulation. The error bars correspond to simulated Poissonian noise ($2 \sigma$).}
	\end{center}
	
\end{figure}

{\it Ab initio} DFT simulations of the dichroic signal were performed using an extension \cite{RuszPRB07} of the WIEN2k package \cite{wien2k} developped for this purpose. In the simulation the effects of thickness, tilt of the incident beam, position of the detector were included as well as the integration over $q_x$ in the range dictated by the use of a circular SEA. Up to 8 beams were used for the calculations of the MDFFs. A comparison with the experiment is given in fig.~\ref{intqyline} for the $L_3$ edge of Cobalt.
The agreement is very good between -0.8 and 0.8 G with some discrepancy appearing at larger scattering angles. This can be due to the faint Bragg spots outside the systematic row (which are neglected in the simulations) and to the fact that the SEA is not exactly in the spectral plane of the energy filter. The error bars correspond to the ($2 \sigma$) Poissonian noise calculated for the theoretical signal using the number of electrons contributing to the signal as determined from the experimental data.

\section{Conclusions}
The strong chiral effect observed in the Co L$_{2,3}$ edge shows that 
\EMCD{} can be measured with reasonable collection time in the TEM. The 
convergence angle of 2~mrad is obviously not detrimental for the 
necessary constant phase shift between the unscattered and the Bragg 
scattered electron waves. The $z$-shift of the specimen allows to control 
the illuminated area, and with optimized conditions a lower limit of 
$\leq$~10~nm appears realistic. This would define the lateral resolution 
in scanning mode. In order to achieve this goal some technical problems 
such as the stability of the beam in the magnetic field, constant $z$-shift 
or decoupling of the scan coordinate from the positioning of the diffraction 
pattern must be solved. The simple concept described above should be an 
incentive for novel dichroic experiments in the TEM. It also shows that \EMCD{} can be complementary and competitive with traditional or new \cite{EisebittNat2004} XMCD techniques.

\newpage
{\bf Acknowledgements}: 

This work was sponsored by the European Union under contract nr. 508971 
(FP6-2003-NEST-A) "Chiraltem". We acknowledge Jo Verbeeck for stimulating discussions.

\bibliographystyle{elsart-num}

\end{document}